\documentclass[conference]{IEEEtran}
\IEEEoverridecommandlockouts
\usepackage{cite}
\usepackage{amsmath,amssymb,amsfonts}
\usepackage{algorithmic}
\usepackage{graphicx}
\usepackage{textcomp}
\usepackage{xcolor}
\usepackage{subcaption}
\usepackage[]{hyperref} 
\def\BibTeX{{\rm B\kern-.05em{\sc i\kern-.025em b}\kern-.08em
    T\kern-.1667em\lower.7ex\hbox{E}\kern-.125emX}}
\begin{document}

\title{SAFE: Semantic Adaptive Feature Extraction with Rate Control for 6G Wireless Communications}

\author{
	\IEEEauthorblockN{
		Yuna Yan\IEEEauthorrefmark{1}, 
		Lixin Li\IEEEauthorrefmark{1}, 
            Xin Zhang\IEEEauthorrefmark{1},
		Wensheng Lin\IEEEauthorrefmark{1}, 
		Wenchi Cheng\IEEEauthorrefmark{2} 
		and Zhu Han\IEEEauthorrefmark{3}} 
	\IEEEauthorblockA{\IEEEauthorrefmark{1}School of Electronics and Information, Northwestern Polytechnical University, Xi’an, China, 710129}
	\IEEEauthorblockA{\IEEEauthorrefmark{2} State Key Laboratory of Integrated Services Networks, Xidian University, Xi'an, China, 710071}
	\IEEEauthorblockA{\IEEEauthorrefmark{3}Department of Electrical and Computer Engineering, University of Houston, Houston, TX, 77004}
    \thanks{This paper has been accepted for publication in IEEE Globecom 2024 workshop.
	}
} 

\maketitle

\begin{abstract}
Most current Deep Learning-based Semantic Communication (DeepSC) systems are designed and trained exclusively for particular single-channel conditions, which restricts their adaptability and overall bandwidth utilization. To address this, we propose an innovative Semantic Adaptive Feature Extraction (SAFE) framework, which significantly improves bandwidth efficiency by allowing users to select different sub-semantic combinations based on their channel conditions. This paper also introduces three advanced learning algorithms to optimize the performance of SAFE framework as a whole. Through a series of simulation experiments, we demonstrate that the SAFE framework can effectively and adaptively extract and transmit semantics under different channel bandwidth conditions, of which effectiveness is verified through objective and subjective quality evaluations.
\end{abstract}


\section{Introduction}
In the context of rapid advancements in areas like smart homes and metaverse \cite{ref1}, there is a rapid increase in the demand for data processing and analysis. This further facilitates a need for the sixth-generation (6G) network to intelligently process and analyze a large amount of data while providing more personalized and intelligent services in a very short period of time for data transmission \cite{Du2019Social}. 
Traditional wireless communication often separates data processing and transmission, ignoring the rich semantic information contained in the data. Deep learning-based semantic communication (DeepSC) system \cite{Lin2024SIC, ref3,  Lin2024SF} only transmits information that the communication parties understand, thereby greatly improving the efficiency of communication resources utilization and meeting the needs of modern society for intelligent communications.

However, most current DeepSC systems are trained under specific channel capacity \cite{ref4,ref5,ref6}, resulting in a fixed semantic code length used under a fixed channel bandwidth, which limits its application scope in practical systems. In some extreme cases, such as heavy network loads or unstable transmission channels, a fixed data compression rate may impose a risk of transmission failure. Especially in terms of image transmission, a poor network environment could severely affect the quality of image transmission, making it impossible for users to receive clear and accurate image information. Therefore, developing flexible and efficient communication systems that can adapt to different network conditions has become an important research topic.

To address these issues, a novel bandwidth and channel quality adaptive scheme called DeepJSCC-l++ is proposed in \cite{ref7}. 
This scheme dynamically assigns weights during network training to map each input image to the desired channel bandwidth. However, this method still relies on training the network under different channel bandwidths as a reference for weight allocation and there hasn't fully eliminated the detrimental effects of transmitting images under fixed channel capacity. In \cite{ref8}, the performance of wireless image transmission under usable bandwidth variation is also discussed. 
However, \cite{ref8} constructs a continuous optimization model where the transmission process spans multiple channel blocks, aiming to reconstruct the image from any previous $ l $ channel blocks. However, this system does not fully consider the diversity of each channel block, leading to information redundancy between the blocks. 
Furthermore, \cite{ref9,ref10,ref11,ref12} explore content-adaptive variable-length transmission strategies, where the encoder determines the occupied channel bandwidth based on the content of the image. To achieve efficient task-oriented wireless communication, \cite{Fu2024Scalable} proposed a content-aware semantic communication framework to identify semantically relevant information for the task, supporting adaptive rate allocation schemes.

In this study, we demonstrate the capability of the Semantic Adaptive Feature Extraction (SAFE) framework and prove its ability to be trained and optimized for any given channel bandwidth. SAFE decomposes the image signal into a series of sub-semantics, where each sub-semantic is transmitted through different channels, effectively mitigating the issues caused by channel capacity limitations. These sub-semantics maintain structural consistency while exhibiting diverse features, ensuring that even when only receiving partial sub-semantics, a reduced yet still acceptable image signal can be decoded at the client. As more available sub-semantics are received, the quality of signal reconstruction improves accordingly. This mechanism not only showcases the flexibility of the Deep Joint Source-Channel Coding (DeepJSCC) but also significantly expands its potential in practical applications.

In summary, the main contributions of this work are as follows:

\begin{itemize}
\item{This paper presents a semantic generator and a semantic synthesizer, we have designed, that can adaptively generate multiple sub-semantics based on the content of the image, allowing users to use channel bandwidth as prior information to select appropriate sub-semantics for transmission and image reconstruction.}
\item{Moreover, we provide three learning algorithms to train our semantic communication network, which decompose the training process of the SAFE network into multiple sub-problems, thereby improving training efficiency and model performance.}
\item{We use the ImageNet100 dataset to train SAFE. Results of a series of tests for evaluating the performance of the proposed framework under different communication channel conditions are presented. Experimental results verify the effectiveness of SAFE in improving the bandwidth efficiency of wireless image transmission and demonstrate its adaptability to different communication channel models.}
\end{itemize}

The remaining part of this paper is organized as follows: Section \ref{sec_2} provides a detailed introduction to the design of the SAFE framework. Section \ref{sec_3} discusses and compares three training strategies for the network. Section \ref{sec_4} presents the results of simulations. Finally, Section \ref{sec_5} concludes this paper with summary of the major findings and some concluding statements.

\section{The Semantic Adaptive Feature Extraction Based Semantic Communication Model}
\label{sec_2}

In this section, we provide a detailed introduction to the proposed SAFE algorithm, including the system model and the SAFE network framework.

\subsection{System Model} 
We consider an image transmission scenario over wireless channels. Let $\boldsymbol{I} \in \mathbb{R}^{C \times H \times W}$ represent the input image, where $ C $, $ H $, and $ W $ denote number of color channels, height, and width of the image, respectively. On the semantic generator side, a semantic mapping (SM) encoder $ g(\boldsymbol{I}, \boldsymbol{\omega}) $ is responsible for transforming the real image $ \boldsymbol{I} $ into a higher-dimensional representation, expanding the image dimension to $ C $ $(C \gg 3)$. Then, it is split into high-dimensional information blocks $ \boldsymbol{I}_i $ of different dimensions, where $i \in\{0,1, \cdots, L\}$ and the dimension of $ \boldsymbol{I}_i $ is $ C_{\mathrm{i}} $ $\left(\sum_{i=0}^L C_{\mathrm{i}}=C\right)$. Here, $ \boldsymbol{\omega} $ represents the parameter set of the semantic encoding network. In addition, the number of high-dimensional information blocks $ L $ can be freely chosen according to practical needs. 

In order to ensure the diversified semantic features and reduce information redundancy, these high-dimensional information blocks $ \boldsymbol{I}_i $ are then passed to the sub-semantic feature extraction (SFE) encoding to obtain sub-semantic blocks $ {\boldsymbol{S}}_{\boldsymbol{i}} $ with a dimension of $ d_{\mathrm{i}} $. The sub-semantic extraction process can be expressed as: ${\boldsymbol{S}}_{\boldsymbol{i}}=c\left(\boldsymbol{I}_{\boldsymbol{i}}, \boldsymbol{\phi}\right)$, where $ c(\cdot) $ is the SFE encoder function, and $ \boldsymbol{\phi} $ is the corresponding parameter set. Afterwards, users select different numbers of sub-semantics ${\boldsymbol{S}}_{\boldsymbol{i}} $, $i \in\{0,1, \cdots, k\}(0<k \leq L)$, based on their own channel conditions, and transmit these extracted sub-semantic blocks separately over different channels. However, noise is inevitably introduced during the channel transmission process. 

\begin{figure}[!t]
	\centering
	\includegraphics[width=3.5in]{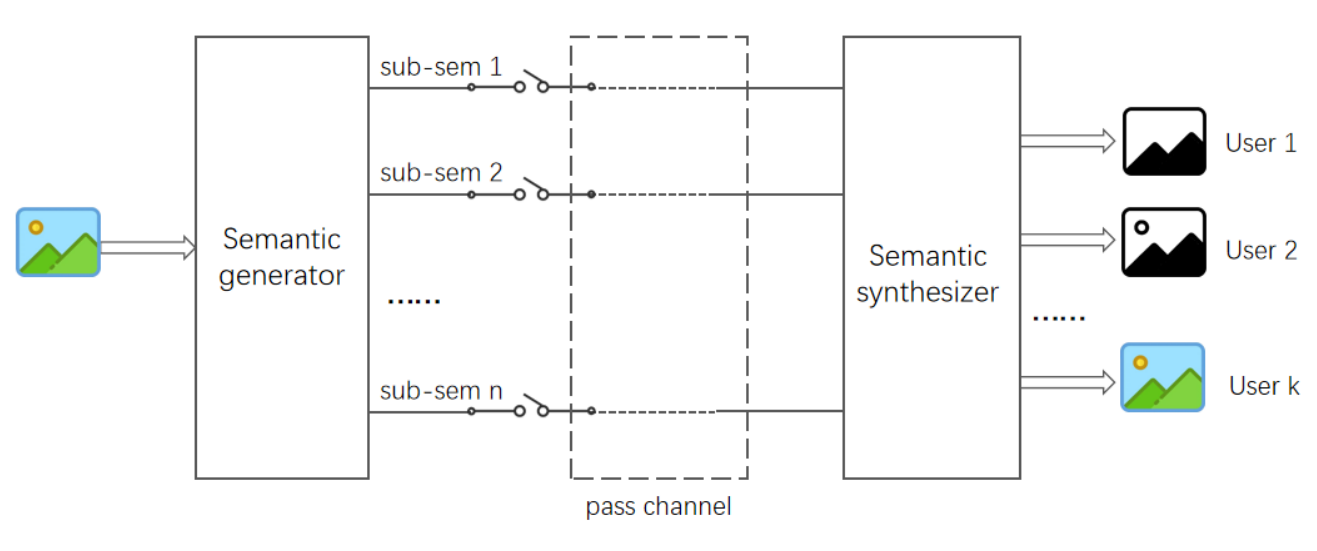}
	\caption{The overall architecture of the proposed SAFE system. }
	\label{fig_1}
\end{figure}

Therefore, the semantic synthesis side uses the sub-semantic feature recovery (SFR) decoder to recover each sub-semantic block $\hat{\boldsymbol{S}}_i$ individually, and reconstructs the original image information $\hat{\boldsymbol{I}}$ by combining these sub-semantics. Even if the user only selects one or a few sub-semantic blocks, or if some sub-semantic blocks are lost, the user can still reconstruct the original image information even if it might be degraded, as shown in Fig. \ref{fig_1}. The more resources the user selects, the better the image restoration quality and the greater the system's reliability. In the most ideal scenario, if the user simultaneously receives all sub-semantic blocks, the original high-quality image can be fully reconstructed through the semantic contraction (SC) decoder, which can be represented by the $ R(\cdot) $ function. The design of this semantic communication system not only enhances the network's adaptability to different channel bandwidths, but also greatly enhances flexibility.

We define the dimension $ n $ of an image as the source bandwidth, and the channel dimension $ k_i $ corresponding to the $ i $-th sub-semantics is the channel bandwidth. The ratio $ {k_i}/n $ is referred to as the bandwidth ratio for the $ i $-th sub-semantic channel. The loss function used for model training is based on the mean squared error (MSE) of $ N $ image samples, given by:
\begin{equation}
	\label{Eq_1}
	\mathrm{MSE}=\frac{1}{n} \sum_{i=1}^n\left(\boldsymbol{I}_j-\hat{\boldsymbol{I}_j}\right)^2,
\end{equation}

where $ \boldsymbol{I}_j $ represents the original image vector of the $ j $-th image sample in the dataset and $ \hat{\boldsymbol{I}_j} $ is the corresponding reconstructed image vector.

\begin{figure*}[!t]
	\centering
	\includegraphics[width=6.5in]{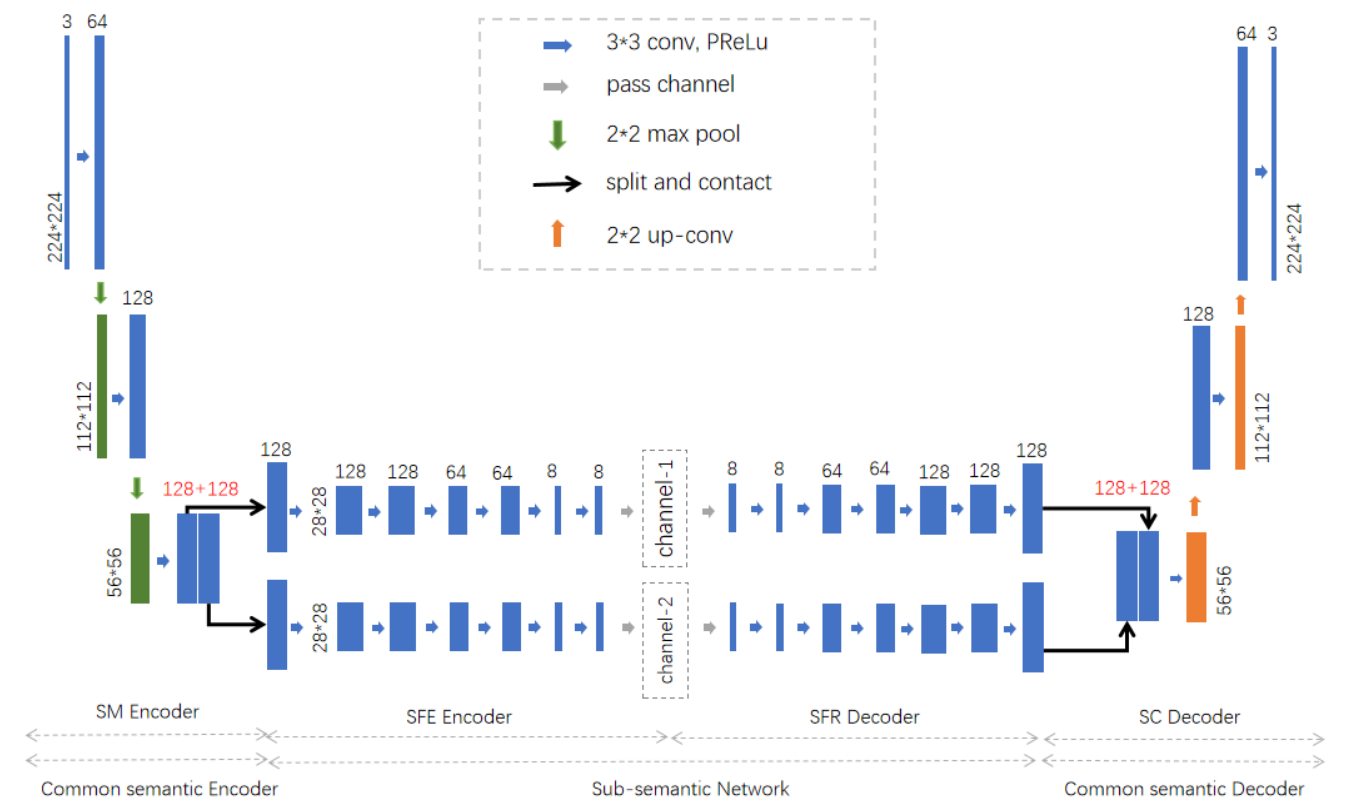}
	\caption{The network framework of the proposed SAFE system. }
	\label{fig_2}
\end{figure*}

\subsection{The Proposed SAFE Network Framework}
We will focus on exploring the semantic communication model of extracting two sub-semantics $ (L = 2) $, The network architecture is shown in Fig. \ref{fig_2}. The design of the semantic communication network with adaptive semantic compression proposed in this paper is inspired by the structure of the original U-Net \cite{ref15} : adopting a 2D U-Net architecture, with the left side responsible for the encoding process (downsampling) and the right side responsible for the decoding process (upsampling). The initial input of the network is an RGB image of size $ 224 \times 224 $, and in the encoding stage, $ 3 \times 3 $ convolutional kernels and $ 2 \times 2 $ Maxpooling are used to reduce the dimensionality of the image. Specifically, after performing convolution, a PReLU activation function is used, followed by a $ 2 \times 2 $ Maxpooling layer with a stride of $ 2 $. In each downsampling step, number of feature channels is twice the original case while the size is a half of it. In the decoding (upsampling) stage, number of feature channels is reduced by half through a $ 3 \times 3 $ transpose convolutional operation, where Upscale represents the upsampling operation, and each transposed convolution is followed by a ReLU activation function. During the decoding process, the network receives the original image and expands the feature map size by a factor of two through the transposed convolution (Deconv), ultimately having output of a processed image with the same size as the initial input. Our design takes into account the independence of the semantic encoder and the decoder in practical applications, thus our network model reduces the connection operation between the left feature map and the right feature map, compared to the original U-Net. Overall, each sub-semantic network contains $ 14 $ convolution layers.

Our research focuses on how to generate multiple sub-semantics based on their correlation and diversified features, in order to perform center reconstruction more effectively. To address this issue, we make the following adjustments to the network:

1) {\em How can we compress semantics as much as possible while accurately restoring the original image?}  This can be achieved by increasing the number of layers in the common semantic network and setting an appropriate number of semantic extraction layers in the common layer. If this part has only a few layers, it is difficult to guarantee exact extraction and transmission of the fundamental semantic information (such as object shape, color, etc.) of the original image. On the contrary, if number of layers in the common semantic network is excessively deep or large, it may hinder the extraction of differentiated semantic features. Therefore, in the common semantic  network layers, we set the depth of the convolutional layers to $ 3 $, ensuring a balance between capturing basic semantic information and preventing potential hindrances to differential feature extraction.

2) {\em How can we ensure that the semantic communication network can extract different differentiated semantic features?} If the number of layers for extracting sub-semantics is too shallow, the preceding layers have already essentially determined the features, and only extracting sub-semantics through a few layers will result in insufficient diversified semantics. Therefore, when extracting sub-semantics, we can increase number of the sub-semantic network layers to gradually compress the semantics. At the same time, we also consider the overall depth of the network to avoid unnecessary complexity. After careful consideration, we set the number of the sub-semantic network layers to $ 4 $ in order to enhance the network's ability to capture a wide range of subtle semantic differences without overwhelming complexity.

Finally, under the condition of equal splitting of sub-semantic dimensions, the bandwidth ratio of the $ i $-th sub-semantic channel is:$  {k_i}/n = (H/8 \times W/8 \times {d_i}) / (H \times W \times 3) = 1/24 $, where $ {d_i}=8 $. Moreover, when all sub-semantics are selected, the total bandwidth ratio is: $ k/n={k_i}/n \times L=1/12 $, with $ L=2 $.

\section{Training Strategies}
\label{sec_3}

\begin{figure} [t!]
	\centering
	\subfloat[Strategy 1]{
		\includegraphics[width=3.5in]{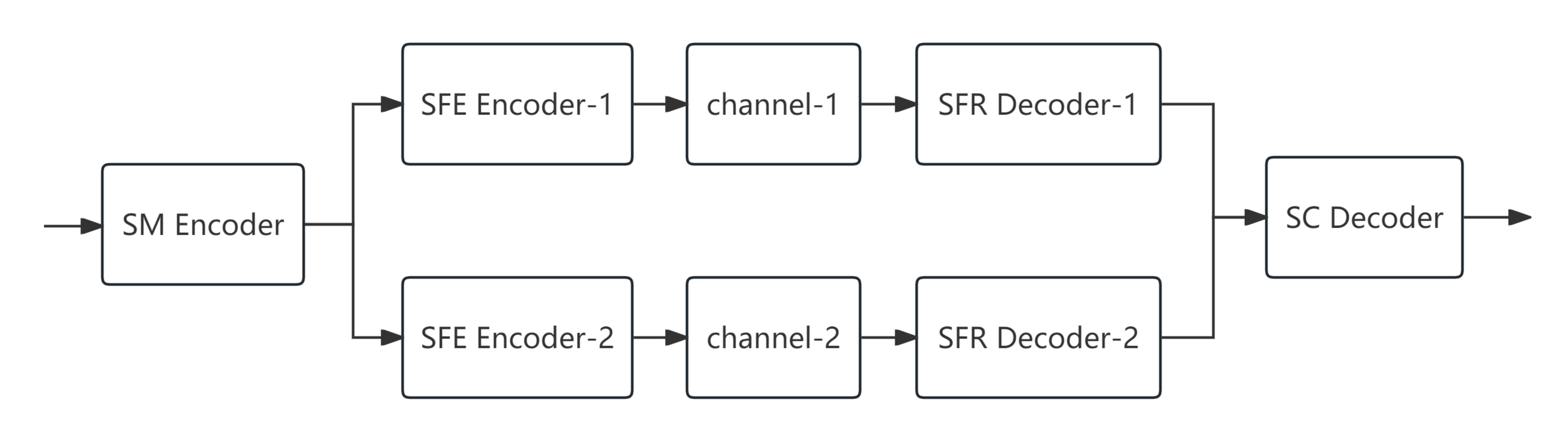}
		\label{3a}}
	\\
	\subfloat[Strategy 2]{
		\includegraphics[width=3.5in]{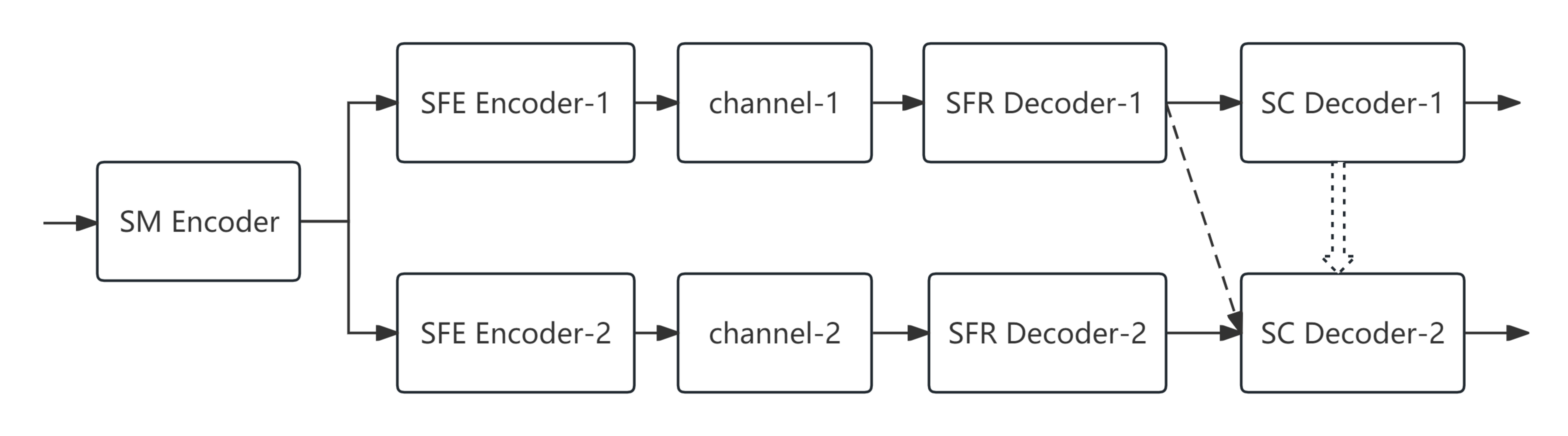}
		\label{3b}}
	\\
	\subfloat[Strategy 3]{
		\includegraphics[width=3.5in]{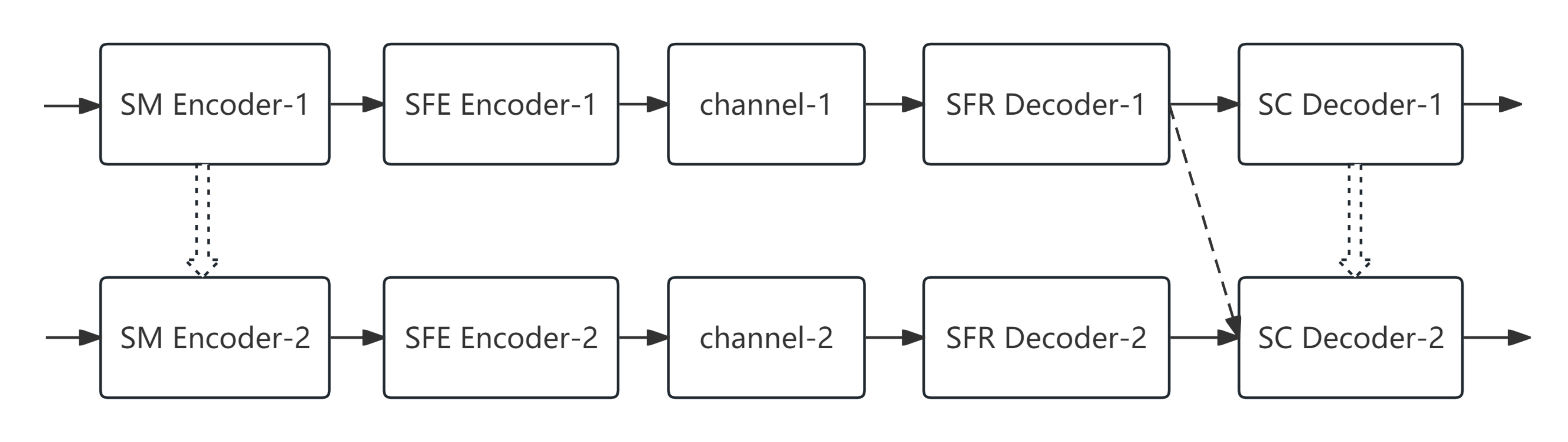}
		\label{3c}}
	\caption{The training flow diagram of three learning algorithms.}
	\label{fig_3} 
\end{figure}

In exploring the network training strategies for sub-semantic information transmission, we propose three different strategies, as follows:

\textbf{Strategy 1}: First, we train a network that focuses on transmitting a single sub-semantic information. At the same time, for the second semantic information, we initialize its input as zero and avoid training the parameters of this sub-semantic network during this stage. After completing the training of the network that transmits only one sub-semantic, we freeze the parameters of this network, thereby shifting our focus to the training of the SFE Encoder-2 and SFR Decoder-2. The advantage of this strategy is that it ensures the performance of the first-level semantic network is not affected, and further improves performance by introducing the second sub-semantic information through gradient descent method. In the worst case, even if all the parameters of the SFE Encoder-2 and SFR Decoder-2 are set to zero, this method can at least guarantee that the loss does not increase.

\textbf{Strategy 2}: Similar to \textbf{Strategy 1}, we first focus on training the first-level semantic network. Then, while keeping the parameters of the SM Encoder, SFE Encoder-2, and SFR Decoder-2 unchanged, we transfer the parameters of the first-level SC Decoder-1 to the second-level SE Decoder, and proceed to train the SFE Encoder-2, SFR Decoder-2, and SC Decoder-2. During the training of the second-level semantic network, we set a lower learning rate for SC Decoder-2 and a relatively higher learning rate for the SFE Encoder-2 and SFR Decoder-2, in order to accelerate the convergence. The characteristic of this strategy is that it does not make interference to  the performance of the first-level semantic network and there is no mutual influence between different decoders. In addition, the parameters of the second-level semantic network can be trained more deeply, and hence the potential performance improvement is more significant. However, the disadvantage of this strategy is that the decoder has more parameters, and the encoder does not have to communicate to negotiate over the parameters to be used (the encoder is not affected).

\textbf{Strategy 3}: This strategy also starts with training the first-level semantic network, in the same way as in \textbf{Strategy 1}. Next, we only freeze the network for the first-level semantics, while transferring the parameters of the first-level SE Encoder-1 and SC Decoder-1 to the second-level SE Encoder and SC Decoder. Subsequently, we train the SE Encoder-2, SFE Encoder-2, SFR Decoder-2, and SC Decoder-2. During this process, we set the learning rate of the SE Encoder-2 and SC Decoder-2 to be relatively low, while setting the learning rate of for the SFE Encoder-2 and SFR Decoder-2 to be relatively high, in order to ensure faster convergence. The advantage of this strategy is that the SE Encoder-2 and SC Decoder-2 are involved in training and can effectively integrate multiple semantic features, potentially demonstrating better performance in handling multiple semantic features. However, this may also lead to a disadvantage: when only the first-level semantic features are transmitted, changes in common semantic layer parameters may cause a degradation in performance. To overcome this problem, we can iteratively train each level of semantic network, the common semantic layer to adapt to different sub-semantic extraction and recovery processes to correct the bias in the common parameters.

It is worth noting that in each strategy, the channel is considered as an untrainable fully-connected layer. In the actual simulation environment, we compare the three strategies described above. The expected outcomes of the three techniques are as follows: 1) \textbf{Strategy 1} may encounter inevitable numerical instability when training the second-level semantic network. 2) Due to the relatively moderate network parameters and the use of the same encoder in \textbf{Strategy 2}, the computational load for training is expected to be lower. Consequently, it can serve as a stable benchmark for performance comparison. 3) As an optimization strategy, \textbf{Strategy 3} demonstrates a higher feasibility. As there is no interference between networks, it is made possible to train each level of the semantic network with optimal parameters. Overall, \textbf{Strategy 2} and \textbf{Strategy 3} can address the learning problem of the SAFE network, but \textbf{Strategy 2} exhibits higher practicality.

\section{Simulation Results}
\label{sec_4}
\subsection{Training Parameters}

\begin{figure} [t!]
	\centering
	\subfloat[AWGN channel]{
		\includegraphics[width=3in]{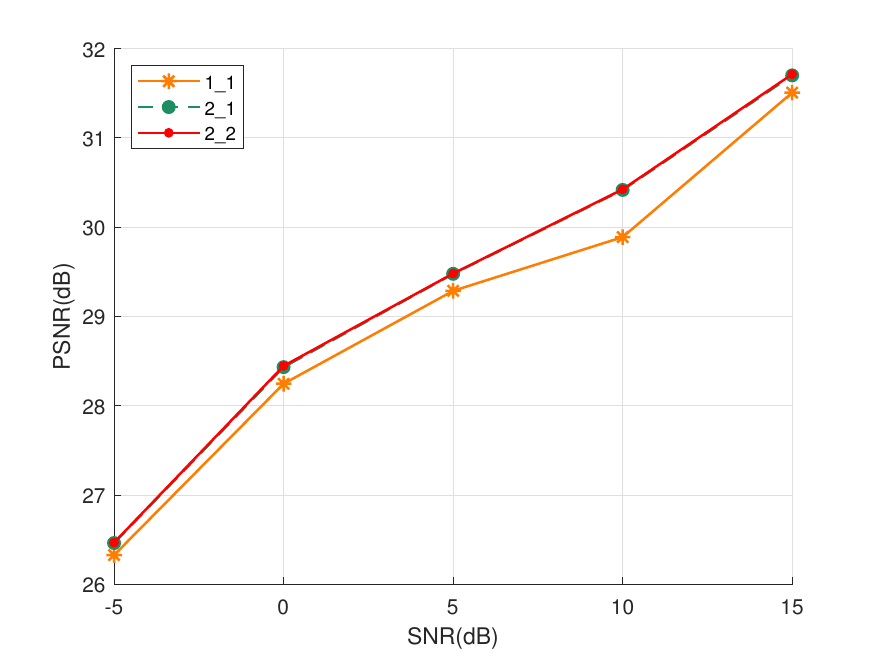}
		\label{4a}}
	\\
	\subfloat[Rayleigh channel]{
		\includegraphics[width=3in]{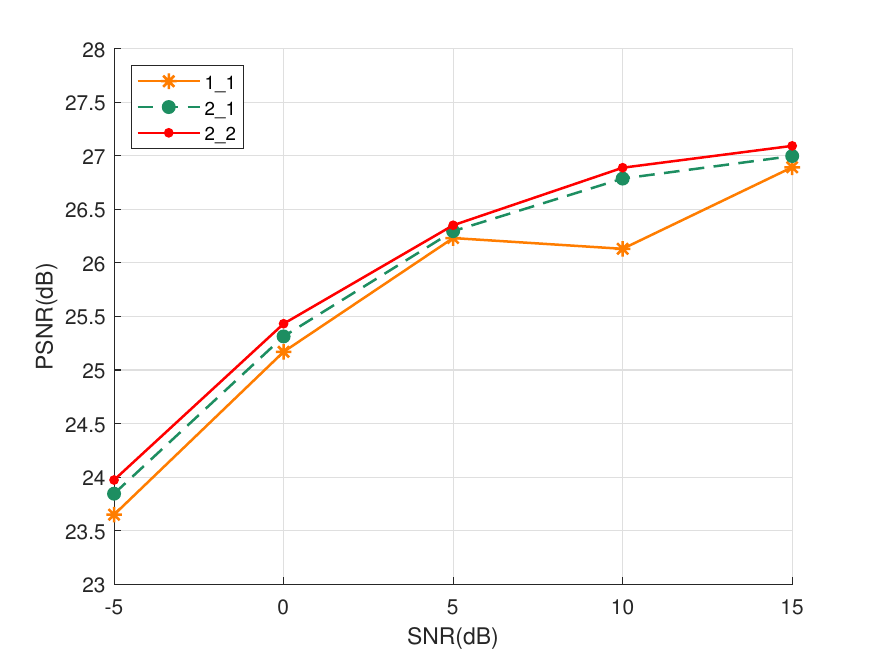}
		\label{4b}}
	\caption{Comparison of PSNR and SNR in Strategy 2 under different channel conditions.}
	\label{fig_4} 
\end{figure}

\begin{figure} [t!]
	\centering
	\subfloat[AWGN channel]{
		\includegraphics[width=3in]{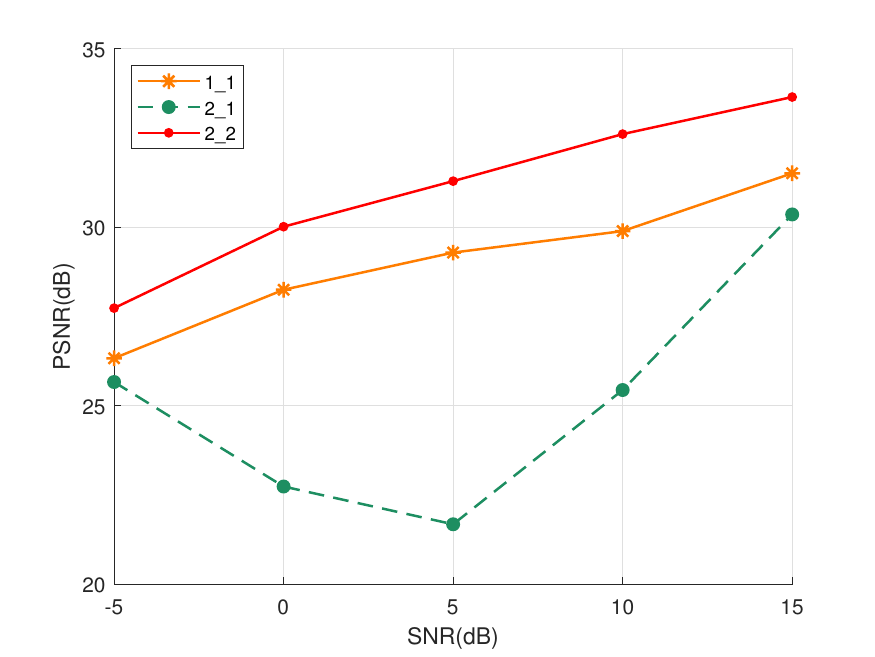}
		\label{5a}}
	\\
	\subfloat[Rayleigh channel]{
		\includegraphics[width=3in]{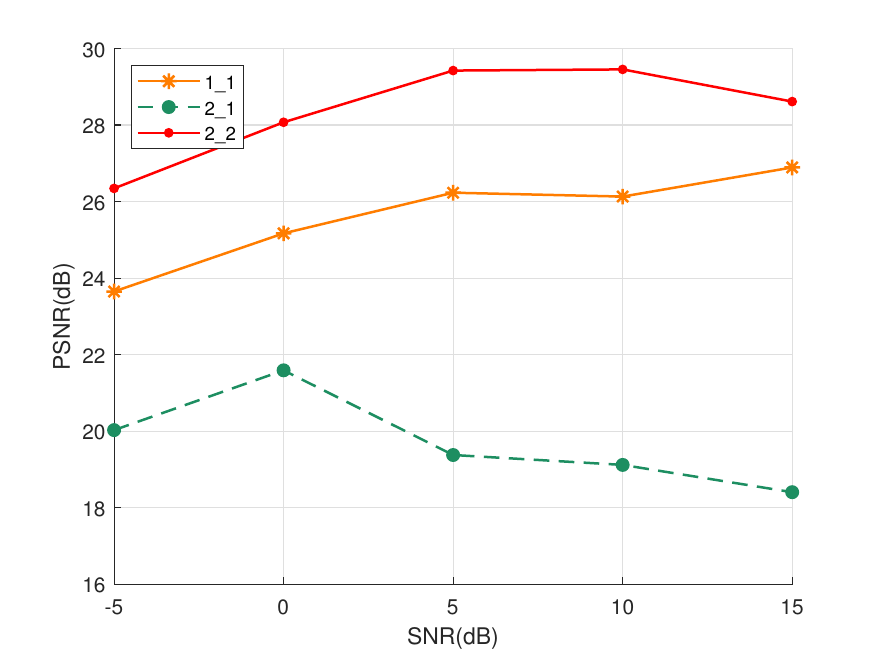}
		\label{5b}}
	\caption{Comparison of PSNR and SNR in Strategy 3 under different channel conditions.}
	\label{fig_5} 
\end{figure}

ImageNet100 is an image dataset consisting of $ 100 $ categories, each containing approximately $ 1000 $ images with a size of $ 224 \times 224 $ pixels. This dataset is a subset selected from the large ImageNet database\cite{ref16}. Based on this dataset, we subdivide it into two parts: the training set and the test set, with the training set containing $ 101,351 $ samples and the test set containing $ 25,338 $ image data. 

Our entire framework is implemented on the PyTorch platform, using the second proposed strategy. We chose to set the training batch size to $ 64 $, and train until there is no improvement for $ 20 $ consecutive epochs. To reduce performance fluctuations, we conducted $ 32 $ repeated experiments to obtain the peak signal-to-noise ratio (PSNR) performance metrics for each image transmission. We utilize the Adam optimizer during the training process and set different learning rates for each sub-semantic network. 

%The specific learning rate Settings are shown in Table \ref{tab:hyperparams}.

%\begin{table}[t!]
%	\caption{Learning rate Settings.}
%	\label{tab:hyperparams}
%	\centering
% \renewcommand{\arraystretch}{1.5}
% \fontsize{3}{4}
% \selectfont
% \resizebox{8cm}{2.5cm}{
	%		\begin{tabular}{c|c}
		%			\hline
		%			Hyper-Parameters          & Value  \\ \hline \hline 
		%			Number of Clients              & 3      \\ \hline
		%			Number of Communication Rounds & 150    \\ \hline
		%			Batch Size          & 32     \\ \hline
		%			Learning Rate       & 0.0001 \\ \hline
		%			Embedding Dimension     & 32     \\ \hline
		%		\end{tabular}
	% }
%\end{table}

\subsection{Experimental Results}

In \textbf{Strategy 2}, as illustrated in Fig. \ref{fig_4}, with increased SNR, there is a clear gradual uptrend in PSNR values across all instances. 
%Furthermore, our method outperforms several representative algorithms in terms of PSNR performance.%
Additionally, compared to the strategy of training and transmitting only one sub-semantic (Train1Trans1), the method for training and transmitting two sub-semantic (Train2Trans2) yields better quality as it includes more semantic information. Simultaneously, the network trained on two sub-semantics also adeptly handles the transmission of one sub-semantic (Train2Trans1). These outcomes convincingly suggest that our strategy effectively addresses the extraction and the restoration for multiple sub-semantics, thereby achieving adaptive compression.

\begin{figure} [t!]
	\centering
	\subfloat[Strategy 2]{
		\includegraphics[width=3.4in]{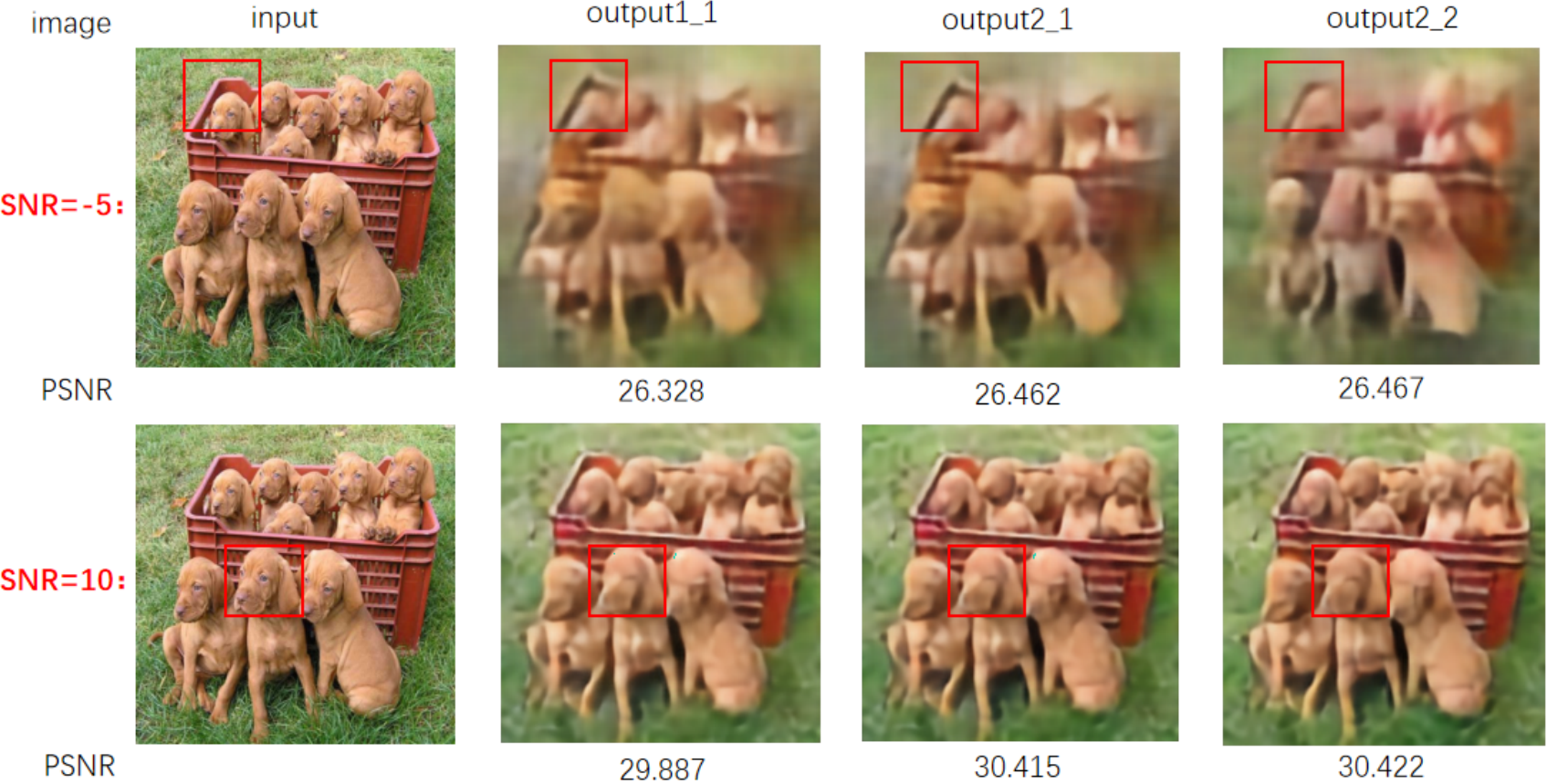}
		\label{6a}}
	\\
	\subfloat[Strategy 3]{
		\includegraphics[width=3.4in]{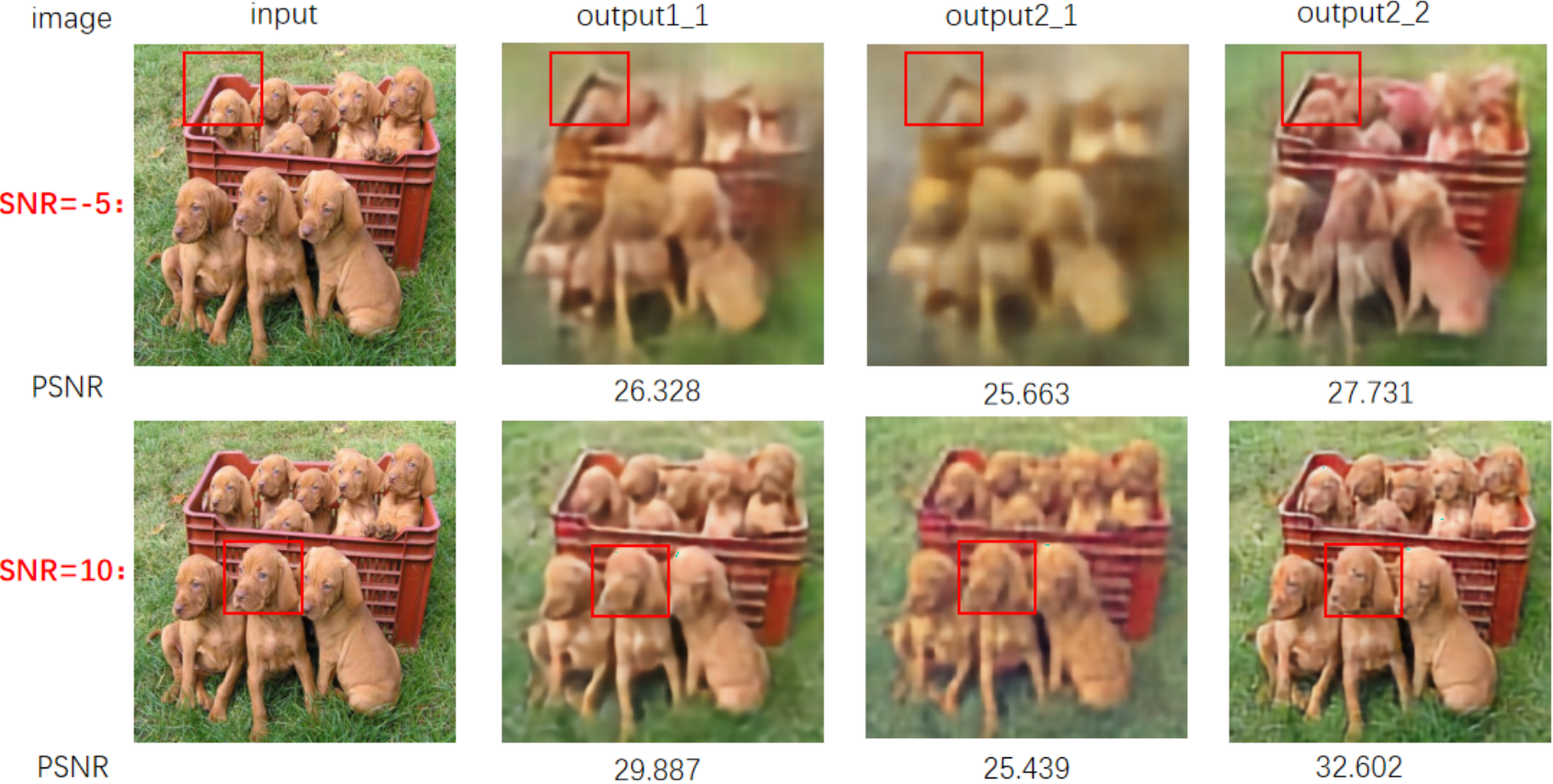}
		\label{6b}}
	\caption{Comparisons of the reconstructed image quality in additive white Gaussian noise channel.}
	\label{fig_6} 
\end{figure}

To further illustrate the optimization effects by the Train2Trans2 method, we conducted \textbf{Strategy 3} and the results presented in Fig. \ref{fig_5}. It is observed that in \textbf{Strategy 3}, the PSNR of the Train2Trans2 method has increased by an average of about 1 to 2 dB (absolute value) compared to \textbf{Strategy 2}. However, different from the \textbf{Strategy 2}, the quality of Train2Trans1 method with \textbf{Strategy 3} is found to be lower than that with Train1Trans1. This gap may be attributed to the involvement of both SM Encoder and SC Decoder in the training process of the second semantic network during transfer training, which significantly altered the optimal parameters in the first semantic network. This also underscores significantly the impact of the effectiveness of SM Encoder in semantic extraction on the quality of semantic communication.

Fig. \ref{fig_6} provides the visual comparison of the different schemes. The outcomes from Train1Trans1 demonstrate that, even under extremely limited bandwidth conditions, the general content of the images, such as shapes and colors, is still preserved. This confirms that our method can effectively compress sub-semantic information while maintaining the fundamental quality of the image. In comparison to Train1Trans1, the images restored by Train2Trans2 exhibit more detailed in terms of edges and textures, exemplifying a significant enhancement in visual quality. As for Train2Trans1, the content of the images remains identifiable, revealing that even a single network is capable of handling the extraction and restoration of multiple sub-semantics. Furthermore, as the SNR increases, the visual quality is improved across all test scenarios. 

\section{Conclusion}
\label{sec_5}
In this paper, we have proposed a SAFE framework that opens up new avenues for wireless image transmission and adaptive rate control based on DeepJSCC. By automatically generating visually similar yet uniquely featured multiple sub-semantics, users can choose the appropriate subset for image restoration under specific bandwidth conditions. We have provided a detailed exposition of the design concept of the SAFE framework and introduced three efficient training strategies. Simulation results demonstrated that, under the same channel conditions and compression ratio, the proposed SAFE model can adapt to various bandwidth configurations, significantly enhancing the versatility and practicality of the model. 

\section*{Acknowledgments}
The authors would like to thank Prof. Tad Matsumoto of JAIST for the suggestions in English revision.

\vspace{12pt}


\begin{thebibliography}{00}

\bibitem{ref1}
Y. Xiao, Q. Du, W. Cheng and W. Zhang, ``Adaptive Sampling and Transmission for Minimizing Age of Information in Metaverse,'' {\em IEEE Journal on Selected Areas in Communications}, vol. 42, no. 3, pp. 588-602, Mar. 2024.

\bibitem{Du2019Social}
 Q. Du, H. Song and X. Zhu, "Social-Feature Enabled Communications Among Devices Toward the Smart IoT Community," \emph{IEEE Communications Magazine}, vol. 57, no. 1, pp. 130-137, January 2019.
 
\bibitem{ref3}
W. Yang, H. Du, Z.Q. Liew, W.Y.B. Lim, Z. Xiong, D. Niyato, X. Chi, X. Shen and C. Miao, ``Semantic Communications for Future Internet: Fundamentals, Applications, and Challenges,'' {\em IEEE Communications Surveys \& Tutorials}, vol. 25, no. 1, pp. 213-250, firstquarter 2023.

\bibitem{Lin2024SF}
W.~Lin, Y.~Yan, L.~Li, Z.~Han, and T.~Matsumoto, ``Semantic-forward relaying:
  {A} novel framework toward {6G} cooperative communications,'' \emph{IEEE
  Communications Letters}, vol.~28, no.~3, pp. 518--522, Mar. 2024.

\bibitem{Lin2024SIC}
W.~Lin, Y.~Yan, L.~Li, Z.~Han, and T.~Matsumoto, ``{SemantIC}: {Semantic}
  interference cancellation toward {6G} wireless communications,'' \emph{IEEE
  Communications Letters}, vol.~28, no.~8, pp. 1810--1814, Aug. 2024.
  
\bibitem{ref4}
E. Bourtsoulatze, D. Burth Kurka and D. Gündüz, ``Deep Joint Source-Channel Coding for Wireless Image Transmission,'' {\em IEEE Transactions on Cognitive Communications and Networking}, vol. 5, no. 3, pp. 567-579, Sep. 2019.

\bibitem{ref5}
H. Xie, Z. Qin, G. Y. Li, and B. H. Juang, ``Deep learning enabled semantic communication systems,'' {\em IEEE Transactions on Signal Processing}, vol. 69, pp. 2663–2675, Apr. 2021. 

\bibitem{ref6}
Z. Weng and Z. Qin, ``Semantic communication systems for speech transmission,'' {\em IEEE Journal on Selected Areas in Communications}, vol. 39, no. 8, pp. 2434–2444, Aug. 2021. 

\bibitem{ref7}
C. Bian, Y. Shao and D. Gündüz ``DeepJSCC-l++: Robust and Bandwidth-Adaptive Wireless Image Transmission,'' {\em 2023 IEEE Global Communications Conference (GLOBECOM)}, Kuala Lumpur, Malaysia, Dec. 2023.

\bibitem{ref8}
D. B. Kurka and D. Gündüz, ``Bandwidth-agile image transmission with deep joint source-channel coding,'' {\em IEEE Transactions on Wireless Communications}, vol. 20, no. 12, pp. 8081–8095, Jun. 2021.

\bibitem{ref9}
W. Cheng, Y. Xiao, S. Zhang and J. Wang, ``Adaptive Finite Blocklength for Ultra-Low Latency in Wireless Communications,'' {\em IEEE Transactions on Wireless Communications}, vol. 21, no. 6, pp. 4450-4463, Jun. 2022.


\bibitem{ref10}
M. Yang and H. -S. Kim, ``Deep Joint Source-Channel Coding for Wireless Image Transmission with Adaptive Rate Control,'' in {\em 2022 IEEE International Conference on Acoustics, Speech and Signal Processing (ICASSP)}, Singapore, May 2022.

\bibitem{ref11}
Y. Zhang, W. Cheng and W. Zhang, ``Multiple Access Integrated Adaptive Finite Blocklength for Ultra-Low Delay in 6G Wireless Networks,'' {\em IEEE Transactions on Wireless Communications}, vol. 23, no. 3, pp. 1670-1683, Mar. 2024.

\bibitem{ref12}
J. Dai, S. Wang, K. Tan, Z. Si, X. Qin, K. Niu, and P. Zhang, ``Nonlinear transform source-channel coding for semantic communications,'' {\em IEEE Journal on Selected Areas in Communications}, vol. 40, no. 8, pp. 2300–2316, Jun. 2022.

\bibitem{Fu2024Scalable}
Y. Fu, W. Cheng, W. Zhang and J. Wang, ``Scalable Extraction Based Semantic Communication for 6G Wireless Networks,'' {\em IEEE Communications Magazine}, vol. 62, no. 7, pp. 96-102, Jul. 2024.     

\bibitem{ref15}
Olaf Ronneberger, Philipp Fischer, and Thomas Brox, ``U-Net: Convolutional Networks for Biomedical Image Segmentation,'' in {\em Medical image computing and computer-assisted intervention–MICCAI 2015: 18th international conference}, Munich, Germany, Oct. 2015.

\bibitem{ref16}
J. Deng, W. Dong, R. Socher, L. -J. Li, Kai Li and Li Fei-Fei, ``Imagenet: A large-scale hierarchical image database,'' in {\em 2009 IEEE Conference on Computer Vision and Pattern Recognition (CVPR)}, Miami, FL, USA, Jun. 2009.

\end{thebibliography}
\end{document}